\documentstyle[a4,12pt,epsf,young]{article}

\begin{document}

\baselineskip=18.8pt plus 0.2pt minus 0.1pt

\def\YGbox{4} 

\font\mybb=msbm10 at 12pt
\def\bb#1{\hbox{\mybb#1}}
\def\Z {\bb{Z}}
\def\R {\bb{R}}
\def\C {\bb{C}}
\def\J {\bb{J}}
\def\I {\bb{I}}
\def\CP {\bb{P}}

\font\mycc=msbm10 at 10pt
\def\cc#1{\hbox{\mycc#1}}
\def\sZ {\cc{Z}}
\def\sR {\cc{R}}

\def\ele{\mathop{\rm ele}\nolimits}
\def\mag{\mathop{\rm mag}\nolimits}
\def\tr{\mathop{\rm tr}\nolimits}
\def\Im{\mathop{\rm Im}\nolimits}
\def\diag{\mathop{\rm diag}\nolimits}
\def\rank{\mathop{\rm rank}\nolimits}
\def\Tr{\mathop{\rm Tr}\nolimits}
\def\mod{\mathop{\rm mod}\nolimits}

\def\sym{\Young{2}\,}
\def\asym{\Young[-0.3]{11}\,}
\def\fnd{\Young{1}\,}

\def\mat#1{\matt[#1]}
\def\matt[#1,#2,#3,#4]{\left(%
\begin{array}{cc} #1 & #2 \\ #3 & #4 \end{array} \right)}

\renewcommand{\theparagraph}{(\roman{paragraph})}
\renewcommand{\thefootnote}{\fnsymbol{footnote}}
\renewcommand{\theequation}{\thesection.\arabic{equation}}
\makeatletter
\@addtoreset{equation}{section}
\@addtoreset{footnote}{page}
\makeatother

\newcommand{\vs}{\vspace*}
\newcommand{\vsx}{\vspace*{1ex}}
\newcommand{\vsh}{\vspace*{0.5ex}}
\newcommand{\vsm}{\vspace*{-1ex}}
\newcommand{\hs}{\hspace*}
\newcommand{\wt}{\widetilde}
\newcommand{\ol}{\overline}
\newcommand{\ul}{\underline}
\newcommand{\ra}{\rightarrow}
\newcommand{\la}{\leftarrow}
\newcommand{\lra}{\leftrightarrow}
\newcommand{\del}{\partial}
\newcommand{\nn}{\nonumber}
\newcommand{\mms}[1]{\makebox[4ex]{$#1$}}
\newcommand{\sq}{\sqrt{2}\,}
\newcommand{\VEV}[1]{\left\langle #1\right\rangle}
\newcommand{\bra}[1]{\left\langle\, #1\,\right|}
\newcommand{\ket}[1]{\left|\, #1\,\right\rangle}
\newcommand{\norm}[1]{\parallel #1\parallel}
\newcommand{\N}{{\cal N}}
\newcommand{\AD}[1]{D$\ol{\mbox{#1}}$}
\newcommand{\oo}{$\ol 9$-$\ol 9$ }
\newcommand{\qo}{$9$-$\ol 9$ }
\newcommand{\oq}{$\ol 9$-$9$ }
\newcommand{\qq}{$9$-$9$ }
\newcommand{\half}{\frac{1}{2}\,}

\newcommand{\NSNS}{{\rm NS\,NS}}
\newcommand{\RR}{{\rm R\,R}}
\newcommand{\beq}{\begin{eqnarray}}
\newcommand{\eeq}{\end{eqnarray}}

\newcommand{\cV}{{\cal V}}


\newcommand{\citen}{\cite}

\begin{titlepage}

\begin{flushright}
YITP-99-25\\
hep-th/9905159
\\
May, 1999
\end{flushright}

\begin{center}
{\large \bf
Anomaly Cancellations in the Type I D9-\AD 9 System\\
\vspace*{1ex}
and the $USp(32)$ String Theory
}

\vspace{1cm}
Shigeki Sugimoto
\footnote{
e-mail: sugimoto@yukawa.kyoto-u.ac.jp
}

\vspace{0.5cm}
{\it
Yukawa Institute for Theoretical Physics,
Kyoto University,\\
Kyoto 606-8502, Japan
}
\end{center}

\vspace{1cm}
\begin{abstract}
We check some consistency conditions for the D9-\AD 9 system in type I
string theory. The gravitational anomaly and gauge anomaly for
 $SO(n)\times SO(m)$ gauge symmetry are shown to be cancelled
when $n-m=32$. In addition, we find that
a string theory with $USp(n)\times USp(m)$ gauge symmetry
also satisfies the anomaly cancellation conditions.
After tachyon condensation, the theory reduces 
to a tachyon-free $USp(32)$ string theory,
though there is no spacetime supersymmetry.
\end{abstract}
\end{titlepage}

\section{Introduction}

To this time, most research on D-branes in string theory
has been carried out in supersymmetric configurations.
The BPS property of branes protects the system from quantum
corrections and provides a nice perspective to go beyond
perturbation in the weakly coupled regime.
In particular, the stability of the BPS D-branes 
is one of the key properties in testing
 various dualities in string theories and
supersymmetric gauge theories.

However, fortunately or unfortunately,
the real world is not supersymmetric, at least in the low energy
scale, and we should engage ourselves in the study of non-supersymmetric
theories sooner or later. Even if one postpones consideration of
the phenomenological
aspects,  there are various interesting features in the
non-BPS configurations of branes,
as should be the case, since most of the states in string theory are
non-BPS.

Recently, the research on the non-BPS configurations of D-branes in
string theory has entered a new stage.
 It was discussed in Refs.~\citen{Sen} and \citen{W} that
 D-branes can be constructed as bound states of
brane-anti-brane systems and several new non-BPS D-branes were
found using this construction.
In this paper, we mainly consider the D9-\AD 9 system in
type I string theory.
 As shown in Ref.~\citen{W}, lower dimensional
D-branes in type I string theory
 can be constructed by arranging non-trivial
Chan-Paton bundles for the D9- and/or \AD 9-branes.
This construction leads to an interpretation in K-theory
and it has been shown that the possible D-branes in type I string theory
can be classified by KO-groups. \cite{W}

We will check some consistency conditions for the D9-\AD 9 system
in this paper. The gauge group of the D9-\AD 9 system
is $SO(n)\times SO(m)$, which is potentially anomalous.
The gravitational and mixed anomalies may also arise in this system.
We will show that these anomalies are all cancelled when
 $n-m=32$. Interestingly, there is another solution of the
anomaly cancellation conditions, which suggests the existence of
a consistent D9-\AD 9 system with $USp(n)\times USp(m)$
gauge symmetry. After tachyon condensation, the theory reduces 
to a tachyon-free $USp(32)$ string theory,
though there is no spacetime supersymmetry.

This paper is organized as follows. In \S \ref{EFT}
we analyze the effective field theory of the D9-\AD 9 system.
We guess the contents of massless fermions in the theory and
check that the gravitational and gauge anomalies as well as
the mixed anomalies are
all cancelled by the Green-Schwarz mechanism.
 In \S \ref{string}, 
we make systematic analyses in perturbative string theory
and give some results that are consistent with the analyses
given in \S \ref{EFT}. 
In \S \ref{Gen}, we
investigate the general formulation of the D9-\AD9 system
and check the anomaly cancellations in stringy calculations.
In \S \ref{USp32}, we discuss some properties of the $USp(32)$
string theory.

\section{Analyses in the effective field theory}
\label{EFT}
\subsection{Green-Schwarz mechanism in the type I D9-\AD9 system}
\label{GSmech}
In this subsection, we determine the massless fermions
 in the type I D9-\AD 9 system,
imposing the Green-Schwarz anomaly cancellation conditions.
Related analyses in the type IIB D9-\AD 9 system are given
in Ref.~\citen{Sre}.

Let us briefly review the Green-Schwarz anomaly cancellation conditions
in type I string theory.\cite{GS}\footnote{We mainly follow the
description in \S 13.5 of Ref.~\citen{GSW}.}
We consider the case in which the gauge group is $SO(n)$.
The gaugino belongs to the adjoint representation
of the gauge group and contributes to the anomaly.
The gravitational anomaly cancellation requires
\begin{eqnarray}
  496=\half n(n-1),
\label{496}
\end{eqnarray}
where the right-hand side is the dimension of the adjoint
representation of $SO(n)$. Equation (\ref{496}) implies the well-known
result $n=32$. Then the rest of the anomaly is
proportional to
\begin{eqnarray}
I_{12}&\propto&-\frac{1}{15}\Tr F^6+\frac{1}{24}\Tr F^4\tr R^2
-\frac{1}{960}\Tr F^2 (4\tr R^4+5(\tr R^2)^2)\nn\\
&&~~+\frac{1}{8}\tr R^2\tr R^4 + \frac{1}{32}(\tr R^2)^3,
\label{anomaly}
\end{eqnarray}
where $F$ is the field strength 2-form of $SO(n)$ and
$R$ is the curvature 2-form.
One of the important identities for the
gauge anomaly cancellation is
\begin{eqnarray}
  \Tr F^6=(n-32)\tr F^6+15\tr F^2\tr F^4.
\label{F6}
\end{eqnarray}
Here we denote traces in the adjoint representation of
$SO(n)$ by the symbol `$\Tr$',
while traces in the fundamental representation are denoted
`$\tr$'. 
In order to cancel the gauge anomaly,
the coefficient of $\tr F^6$ on the right-hand side of
(\ref{F6}) must vanish. This condition is also satisfied
when $n=32$. Then all of the terms in (\ref{anomaly})
are cancelled by the counterterm
\begin{eqnarray}
\Delta\Gamma=\int B X_8-\left(\frac{2}{3}+\alpha\right)
\int(\omega_{3L}-\omega_{3Y})X_7,
\label{DG}
\end{eqnarray}
where $B$ is the 2-form field and $\alpha$ is an 
adjustable parameter. The quantity $X_8$ is given as
\begin{eqnarray}
X_8&=&\tr F^4-\frac{1}{8}\tr F^2\tr R^2
+\frac{1}{8}\tr R^4+\frac{1}{32}(\tr R^2)^2,
\end{eqnarray}
and $\omega_{3L}$, $\omega_{3Y}$ and $X_7$ are defined by
\begin{eqnarray}
\tr R^2=d\omega_{3L},\qquad \tr F^2=d\omega_{3Y}, \qquad X_8=dX_7,
\end{eqnarray}
modulo exact forms.
The counterterm (\ref{DG}) induces an anomaly of the form
\begin{eqnarray}
I_{12}\propto(\tr R^2-\tr F^2)X_8,
\end{eqnarray}
which exactly cancels the anomaly (\ref{anomaly}).

Now, consider type I string theory with $n$ D9-branes
and $m$ \AD 9-branes. The gauge group is
$SO(n)\times SO(m)$.
The ordinary type I string theory
corresponds to the case in which $n=32$ and $m=0$.
Since several observations
suggest that we can create or annihilate Dp-\AD p pairs without
changing the physical context,\cite{Sen,W}
 it is natural to assume that the coefficients of the
anomalies only depend on $n-m$.
Then the condition corresponding to (\ref{496}) becomes
\begin{eqnarray}
  496&=&\half (n-m)(n-m-1)\\
  &=&\half n(n-1)+\half m(m+1) -mn,
\label{mn}
\end{eqnarray}
which implies $n - m = 32$.
Equation (\ref{mn}) suggests that the \qq fermions $\lambda$
and the \oo fermions $\wt\lambda$ are positive chirality
spinors, which belong to the adjoint representation
of $SO(n)$ and the second rank symmetric 
tensor representation of $SO(m)$, respectively,
while the \qo or \oq fermions $\psi$ are
negative chirality spinors, which belong to the bifundamental
representation of $SO(n)\times SO(m)$ (Table \ref{table1}).
\begin{center}
\refstepcounter{table}
\label{table1}
\footnotesize
{\bf Table \thetable:} Fermions in the D9-\AD 9 system.
\vsx
$$
\begin{array}{c|c|c|c}
\hline \hline
\mbox{string}&\mbox{fermion}
& \mbox{rep. of $SO(n)\times SO(m)$} & \mbox{chirality}\\
\hline
\mbox{\qq string}&\lambda&(\asym,1)&+\\
\mbox{\oo string}&\wt\lambda&(1,\sym)&+\\
\mbox{\qo$\!\!$,\,\oq string}&\psi&(\fnd,\fnd)&-\\
\hline
\end{array}
\nn
$$
\end{center}
\vsh

Let us show that all the anomalies are cancelled when we choose
these fermions. We denote the field strength 2-forms
of $SO(n)$ and $SO(m)$ gauge fields by $F_1$ and $F_2$,
respectively.
 We have the identities
\begin{eqnarray}
  \Tr_1 F_1^6&=&(n-32)\tr_1 F_1^6+15\tr_1 F_1^2\tr_1 F_1^4,
\label{F6-1}\\
  \Tr_2 F_2^6&=&(m+32)\tr_2 F_2^6+15\tr_2 F_2^2\tr_2 F_2^4,
\label{F6-2}\\
  \Tr_1 F_1^4&=&(n-8)\tr_1 F_1^4+3(\tr_1 F_1^2)^2,\\
  \Tr_2 F_2^4&=&(m+8)\tr_2 F_2^4+3(\tr_2 F_2^2)^2,\\
  \Tr_1 F_1^2&=&(n-2)\tr_1 F_1^2,\\
  \Tr_2 F_2^2&=&(m+2)\tr_2 F_2^2,
\label{F2}
\end{eqnarray}
where we label traces in the representations of
$SO(n)$ and $SO(m)$ by the
subscripts 1 and 2, respectively. We denote traces in the
adjoint representation of $SO(n)$ by `$\Tr_1$',
while
`$\Tr_2$' represents the traces in the second rank symmetric
tensor representation of $SO(m)$.
Traces in the
fundamental representations of $SO(n)$ and $SO(m)$
 are denoted `$\tr_i$' ($i=1,2$).
Collecting all the contributions of the fermions
$\lambda,~\wt\lambda$ and $\psi$,
the term $\Tr F^6$ in (\ref{anomaly}) is replaced by
\begin{eqnarray}
\Tr F^6&\ra& \Tr_1 F_1^6+\Tr_2 F_2^6
  -m \tr_1 F_1^6-n \tr_2 F_2^6\nn\\
&&  -\left(6\atop 2\right)(\tr_1F_1^2\tr_2F_2^4+\tr_1F_1^4\tr_2F_2^2),
\label{F1F2}
\end{eqnarray}
where the first two terms in (\ref{F1F2}) are the contributions of
the \qq fermion $\lambda$ and the \oo fermion $\wt\lambda$,
while the other terms are the contributions of the \qo fermion $\psi$.
Using the identities (\ref{F6-1}) and (\ref{F6-2}),
it is easy to check that the coefficients of 
$\tr_1 F_1^6$ and $\tr_2 F_2^6$ in (\ref{F1F2})
vanish if and only if $n-m=32$.
Similarly, $\Tr F^4$ and $\Tr F^2$ in (\ref{anomaly}) are replaced by
\begin{eqnarray}
\Tr F^4&\ra& \Tr_1 F_1^4+\Tr_2 F_2^4
  -m \tr_1 F_1^4-n \tr_2 F_2^4\nn\\
&&  -\left(4\atop 2\right)\tr_1F_1^2\tr_2F_2^2,\\
\Tr F^2&\ra& \Tr_1 F_1^2+\Tr_2 F_2^2
  -m \tr_1 F_1^2-n \tr_2 F_2^2.
\end{eqnarray}
Then, using the identities (\ref{F6-1})--(\ref{F2}),
the anomaly (\ref{anomaly}) can be written as
\beq
I_{12}&\propto&(\tr R^2-\tr_1F_1^2+\tr_2F_2^2)X'_8,
\eeq
\beq
X'_8&=&\tr_1F_1^4-\tr_2F_2^4-\frac{1}{8}(\tr_1 F_1^2-\tr_2 F_2^2)\tr R^2\nn\\
&&+\frac{1}{8}\tr R^4+\frac{1}{32}(\tr R^2)^2.
\eeq
This anomaly can be cancelled by the counterterm
\begin{eqnarray}
\Delta\Gamma=\int B X'_8-\left(\frac{2}{3}+\alpha\right)
\int(\omega_{3L}-\omega_{3Y_1}+\omega_{3Y_2})X'_7,
\end{eqnarray}
where we have defined $\tr_iF^2=d\omega_{3Y_i}$ and $X'_8=dX'_7$.
The gauge invariant combination of the field strength of the 2-form field
$B$ is now
\beq
H=dB+\omega_{3L}-\omega_{3Y_1}+\omega_{3Y_2}.
\eeq

\subsection{Coupling to the tachyon fields}
\label{condense}
There are tachyon fields in the D9-\AD 9 system.
After tachyon condensation, the D9-\AD 9 brane pairs are
expected to vanish, and the field content of the theory
turns out to be the same as that of type I $SO(32)$ string theory.
Let us show that the fermion contents given in Table \ref{table1}
are also suitable to explain the brane-anti-brane pair annihilation.


The tachyon fields $T_{i\bar j}$ belong to the bifundamental
representation of $SO(n)\times SO(m)$. 
We denote $i,~j=1,\cdots,n$ as the vector indices
of $SO(n)$ and $\bar i,~\bar j=1,\cdots,m$ as the vector indices of
$SO(m)$. The following Yukawa interactions are consistent with the
symmetry:
\beq
{\cal L}_Y\sim\,
\ol\psi^{i\bar i}\lambda^{ij}T_{j\bar i},
+\ol\psi^{i\bar i}\wt\lambda^{\bar i\bar j}T_{i\bar j},
\label{yukawa}
\eeq
where $\ol\psi=\psi^T\Gamma^0$.

The tachyon VEV is of the form
\beq
&&~~\longleftarrow ~~~~~n~~~~~ \longrightarrow\nn\\
(T_{i\bar j})&=&\left(
\begin{array}{cccccc}
\hspace{5ex}&{*}\\
&&\ddots\\
&&&{*}
\end{array}
\right)
\begin{array}{c}
\uparrow\\
m\\
\downarrow
\end{array}
\label{VEV}\\
&&~~\la 32\ra\la~m~\ra\nn
\eeq
up to gauge symmetry. It is plausible to assume that the symbols
${*}$ in
(\ref{VEV}) are all non-zero, although
we do not know the precise form
of the tachyon potential. Then, the Yukawa terms (\ref{yukawa})
will induce mass terms for $\psi$, $\wt\lambda$ and $\lambda^{ij}$
($i>32$ or $j>32$). 
The number of the components of
$\wt\lambda$ and $\lambda^{ij}$ ($i>32$ or $j>32$) are
$\half m(m+1)$ and $\half m(m-1)+32 m$, respectively,
and the sum is just enough to be paired with
$\psi$, which has $mn=m^2+32m$ components.
 As a result, the massless
components of the fermions after the tachyon condensation
 are $\lambda^{ij}$ ($i,~j=1,\cdots,32$),
 which belong to the adjoint representation of the unbroken
$SO(32)$ gauge group, as expected.

\vspace*{2mm}

\section{Analyses in string theory}
\label{string}

\vspace*{2mm}
 
\subsection{Physical states in the type I D9-\AD9 system}
\label{Physstat}
The \AD9-brane is a 9-brane with $-1$ units of
R-R charge. It can be obtained by flipping the sign
of the R-R charge of a D9-brane. Using this fact,
we can compute the vacuum amplitudes in the D9-\AD9 system,
from which the physical spectrum can be extracted,
as discussed in Ref.~\citen{Sen}. 
As a preliminary step, let us first
collect here the one-loop vacuum amplitudes for the \qq strings.
\cite{Pol}
There are the contributions from  the  NS sector and the Ramond sector,
which are denoted as $Z^{\rm NS}$ and $Z^{\rm R}$.
We decompose these terms 
into the contributions of NS-NS exchange and R-R exchange
 in the closed string channel, denoted $Z_{\NSNS}$ and
$Z_{\RR}$. 
We also label the contributions from the cylinder diagram and 
the M\"obius strip diagram for the amplitudes
with superscripts as $Z^{\rm (C_2)}$ and $Z^{\rm (M_2)}$:
\beq
Z_{9\hbox{-}9}&=&Z_{9\hbox{-}9}^{\rm NS}+Z_{9\hbox{-}9}^{\rm R},\\
Z_{9\hbox{-}9}^{\rm NS}&=&Z^{\rm NS~(C_2)}_{\NSNS}+
Z^{\rm NS~(M_2)}_{\NSNS}+
Z^{\rm NS~(C_2)}_{\RR}+
Z^{\rm NS~(M_2)}_{\RR},\\
Z_{9\hbox{-}9}^{\rm R}&=&Z^{\rm R~(C_2)}_{\NSNS}+
Z^{\rm R~(M_2)}_{\NSNS}+
Z^{\rm R~(C_2)}_{\RR}+
Z^{\rm R~(M_2)}_{\RR},
\eeq
\beq
Z^{\rm NS~(C_2)}_{\NSNS}&=&\int^\infty_0\frac{dl}{8l}\Tr_{\rm NS}
\left(\exp(-Hl)\right),\\
&=&in^2 V_{10}\int^\infty_0\frac{dt}{4t}(8\pi^2\alpha't)^{-5}
\eta(it)^{-8}Z^0_0(it)^4,\\
&=&in^2 V_{10}\int^\infty_0\frac{dt}{4t}(8\pi^2\alpha't)^{-5}
(q^{-1/2}+8+O(q^{1/2})),\\
Z^{\rm NS~(M_2)}_{\NSNS}&=&\int^\infty_0\frac{dl}{8l}\Tr_{\rm NS}
\left(\Omega(1+(-1)^F)\exp(-Hl)\right),\\
&=&- inV_{10}\int^\infty_0\frac{dt}{4t}(8\pi^2\alpha't)^{-5}
\frac{Z^0_1(2it)^4Z^1_0(2it)^4}{\eta(2it)^{8}Z^0_0(2it)^4},\\
&=&- in V_{10}\int^\infty_0\frac{dt}{4t}(8\pi^2\alpha't)^{-5}
(16+O(q)),\\
Z^{\rm NS~(C_2)}_{\RR}&=&\int^\infty_0\frac{dl}{8l}\Tr_{\rm NS}\left(
(-1)^F\exp(-Hl)\right),\\
&=&in^2 V_{10}\int^\infty_0\frac{dt}{4t}(8\pi^2\alpha't)^{-5}
\eta(it)^{-8}\left(-Z^0_1(it)^4\right),\\
&=&in^2 V_{10}\int^\infty_0\frac{dt}{4t}(8\pi^2\alpha't)^{-5}
(-q^{-1/2}+8+O(q^{1/2})),\\
Z^{\rm NS~(M_2)}_{\RR}&=&0,\\
Z^{\rm R~(C_2)}_{\NSNS}&=&-\int^\infty_0\frac{dl}{8l}\Tr_{\rm R}\left(
\exp(-Hl)\right),\\
&=&in^2 V_{10}\int^\infty_0\frac{dt}{4t}(8\pi^2\alpha't)^{-5}
\eta(it)^{-8}\left(-Z^1_0(it)^4\right),\\
&=&in^2 V_{10}\int^\infty_0\frac{dt}{4t}(8\pi^2\alpha't)^{-5}
(-16+O(q)),\\
Z^{\rm R~(M_2)}_{\NSNS}&=&0,\\
Z^{\rm R~(C_2)}_{\RR}&=&-\int^\infty_0\frac{dl}{8l}\Tr_{\rm R}
\left((-1)^F\exp(-Hl)\right)=0,\\
Z^{\rm R~(M_2)}_{\RR}&=&-\int^\infty_0\frac{dl}{8l}\Tr_{\rm R}\left(
\Omega(1+(-1)^F)\exp(-Hl)\right),\\
&=&+ inV_{10}\int^\infty_0\frac{dt}{4t}(8\pi^2\alpha't)^{-5}
\frac{Z^0_1(2it)^4Z^1_0(2it)^4}{\eta(2it)^{8}Z^0_0(2it)^4},\\
&=&+ in V_{10}\int^\infty_0\frac{dt}{4t}(8\pi^2\alpha't)^{-5}
(16+O(q)),
\eeq
where $q=e^{-2\pi t}$, and
\beq
Z^0_0(it)&=&q^{-1/24}\prod_{m=1}^\infty(1+q^{m-1/2})^2,\\
Z^0_1(it)&=&q^{-1/24}\prod_{m=1}^\infty(1-q^{m-1/2})^2,\\
Z^1_0(it)&=&2q^{1/12}\prod_{m=1}^\infty(1+q^m)^2.
\eeq


Then we have
\beq
Z^{\rm NS}_{9\hbox{-}9}&=&iV_{10}\int^\infty_0\frac{dt}{t}(8\pi^2\alpha't)^{-5}
\left(8\cdot\frac{1}{2}n(n-1)+O(q)\right),\\
Z^{\rm R}_{9\hbox{-}9}&=&iV_{10}\int^\infty_0\frac{dt}{t}(8\pi^2\alpha't)^{-5}
\left(-8\cdot\frac{1}{2}n(n-1)+O(q)\right).
\eeq

The contributions of the \oo strings can be obtained
by replacing $n$ by $m$ and flipping the sign of the
R-R charge of the D9-brane. In the one-loop vacuum amplitudes,
we should flip the sign of $Z_{\RR}^{\rm R~(M_2)}$,
since the contribution of the R-R exchange
between the D9-brane boundary state and the cross-cap state
is proportional to the R-R charge of the D9-brane.
Thus we obtain
\beq
Z^{\rm NS}_{\ol 9\hbox{-}\ol 9}
&=&iV_{10}\int^\infty_0\frac{dt}{t}(8\pi^2\alpha't)^{-5}
\left(8\cdot\frac{1}{2}m(m-1)+O(q)\right),
\label{gauge}\\
Z^{\rm R}_{\ol 9\hbox{-}\ol 9}
&=&iV_{10}\int^\infty_0\frac{dt}{t}(8\pi^2\alpha't)^{-5}
\left(-8\cdot\frac{1}{2}m(m+1)+O(q)\right).
\label{fermi}
\eeq
Equation (\ref{gauge}) is the contribution of the $SO(m)$ gauge fields,
and (\ref{fermi}) is the contribution of the massless fermions.
Equation (\ref{fermi}) suggests that the fermions belong to
a second rank symmetric tensor representation of the gauge group
$SO(m)$, as discussed in the last section.
 To confirm this observation in a more systematic way,
note that we have taken an opposite $\Omega$ projection
in the Ramond sector, which corresponds to
the sign flip $Z_{\RR}^{\rm R~(M_2)}\ra-Z_{\RR}^{\rm R~(M_2)}$.
The action of $\Omega$ on the massless fermions is
\beq
\Omega\ket{s\,;ij}(\wt\lambda_s)_{ij}=
-\ket{s\,;ij}\gamma_{jj'}^{-1}(\wt\lambda_s)_{j'i'}\gamma_{i'i},
\label{proj}
\eeq
where $\gamma_{ij}=\delta_{ij}$ for $SO(m)$ theory and
$\gamma_{ij}=i\J_{ij}$ for $USp(m)$ theory. For the Ramond sector of
the \oo string,
we take the states with $\Omega=-1$ as the physical states,
and thus (\ref{proj}) implies
\beq
\wt\lambda=\wt\lambda^{T}.
\label{psisym}
\eeq

The one-loop vacuum amplitudes for \oq and \qo strings are
obtained by replacing $n^2$ with $2nm$ in the cylinder diagrams
and flipping the sign of
the contributions of the R-R exchange between
the D9-brane and \AD 9-brane boundary states.
Then we have
\beq
Z^{\rm NS}_{9\hbox{-}\ol 9,\ol 9\hbox{-}9}
&=&iV_{10}\int^\infty_0\frac{dt}{t}(8\pi^2\alpha't)^{-5}
\left(mn\cdot q^{-1/2}+O(q^{1/2})\right),
\label{tachyon}\\
Z^{\rm R}_{\ol 9\hbox{-}\ol 9,\ol 9\hbox{-}9}
&=&iV_{10}\int^\infty_0\frac{dt}{t}(8\pi^2\alpha't)^{-5}
\left(-8\cdot mn+O(q)\right).
\label{fermi2}
\eeq

{}From (\ref{tachyon}), we conclude that there are $mn$ tachyon fields
in the open string channel. Equation (\ref{fermi2}) is the contribution
of $mn$ massless fermions. It may be useful to write down
the physical state conditions for \qo and \oq strings,
as given in Ref.~\citen{Sen}.
The sign flip $Z_{\RR}^{\rm (C_2)}\ra -Z_{\RR}^{\rm (C_2)}$
means that we have taken an opposite GSO projection
for \qo and \oq strings, i.e.,  $(-1)^F=-1$.
Since we have
\beq
\Omega^2=(-1)^F,
\eeq
in the NS sector, we should take the states
with $\Omega=\pm i$ as physical states in the NS sector. 
We choose the convention $\Omega=i$ in the following.
Since the action of $\Omega$ on the NS ground state is given by
\beq
\Omega\ket{0;i\bar j}_{\rm NS}=i\,(\gamma_{\bar 9})_{\bar j\bar j'}
\ket{0;\bar j' i'}_{\rm NS}(\gamma_9^{-1})_{i'i},
\label{OmeTach}
\eeq
where $i,~i'=1,\cdots,n$ are D9-brane Chan-Paton indices and
$\bar j,~\bar j'=1,\cdots,m$ are \AD 9-brane Chan-Paton indices.
In the present case, since the gauge group is the $SO$ group, we have
$(\gamma_9)_{i'i}=\delta_{i'i}$ and
$(\gamma_{\bar 9})_{\bar j\bar j'}=\delta_{\bar j\bar j'}$.
The tachyon field created by the \oq and \qo strings are
combined as
\beq
T=\mat{,T_{\bar i j},T_{i\bar j},},
\eeq
which is an $(n+m)\times (n+m)$ Hermitian matrix.
Imposing the physical state condition $\Omega=i$, we have
\beq
T^{T}=\gamma^{-1}T\gamma,
\eeq
where
\beq
\gamma=\mat{\gamma_9,,,\gamma_{\bar 9}},
\eeq
leaving $nm$ components as the physical tachyon.

For the Ramond sector ground states, the operator
$(-1)^F$ is equivalent
to the chirality operator $\Gamma$. Thus, if we take
an opposite GSO projection for \qo and \oq strings,
the chirality of the fermions created by these string
is  opposite to the chirality of \qq and \oo fermions.
This result is consistent with the anomaly cancellation conditions
discussed in the previous section.

\subsection{The R-R tadpole cancellation in the type I 
D9-\AD9 system}
\label{RRtad}
The R-R tadpole cancellation is one of the most important
constraints in a consistent string theory.
In the D9-\AD 9 system, the R-R tadpole cancellation
requires the condition $n-m=32$, which we encountered
in the previous section, (\ref{mn}). Though
this condition can be easily understood by counting
the R-R charges of D9-branes, \AD 9-branes and an O$9^-$-plane,
it would be instructive to demonstrate the explicit calculation
in our framework. 

The divergences due to the R-R tadpole can be extracted by
the modular transformation in
one-loop vacuum amplitudes $Z_{\RR}$.
Using the identities
\beq
\eta(it)=t^{-1/2}\eta(i/t),\qquad Z^\alpha_\beta(it)=Z^\beta_{\alpha}(i/t),
\eeq
and defining $s=\pi/t$ for the cylinder and
$s=\pi/4t$ for the M\"obius strip,
we have
\beq
Z_{\RR}^{\rm NS~(C_2)}&=&n^2\cdot\frac{iV_{10}}{8\pi(8\pi^2\alpha')^5}
\int^\infty_0 ds \left(-16+O(e^{-2s})\right),\\
Z_{\RR}^{\rm R~(M_2)}&=&+2^6n\cdot\frac{iV_{10}}{8\pi(8\pi^2\alpha')^5}
\int^\infty_0 ds \left(16+O(e^{-2s})\right).
\eeq
There are also the contributions from the Klein bottle diagram,
\beq
Z_{\RR}^{\rm (K_2)}=2^{10}\cdot\frac{iV_{10}}{8\pi(8\pi^2\alpha')^5}
\int^\infty_0 ds \left(-16+O(e^{-2s})\right).
\eeq
The contributions from the \oq$\!\!$,\
\qo and \oo strings can be
obtained similarly.
The results are summarized in Table~\ref{table2},
where we have suppressed the divergent factor
\begin{eqnarray}
\frac{iV_{10}}{8\pi(8\pi^2\alpha')^5}
\int^\infty_0 ds \left(-16+O(e^{-2s})\right).
\end{eqnarray}
\begin{center}
\refstepcounter{table}
\label{table2}
\footnotesize
{\bf Table \thetable:} The divergences due to R-R tadpole.
\vsx
\beq
\begin{array}{c|c|c|c}
\hline\hline
\mbox{string}&Z_{\RR}^{\rm (C_2)}
&Z_{\RR}^{\rm (M_2)}&Z_{\RR}^{\rm (K_2)}\\
\hline
\mbox{\qq string}&n^2&-2^6n&\\
\mbox{\oo string}&m^2&+2^6m&\\
\mbox{\qo$\!\!$,\,\oq string}&-2mn&&\\
\mbox{closed string}&&&2^{10}\\
\hline
\end{array}
\nn
\eeq
\end{center}
The total contribution is
\beq
n^2+m^2-2mn-2^6n+2^6m+2^{10}=(n-m-32)^2,
\eeq
which is cancelled if and only if $n-m=32$,
as expected.

\subsection{The $USp(n)\times USp(m)$ theory}

The analyses in \S\ref{EFT}
can also be applied to the case in which the gauge group is
the symplectic group. 
The \qq strings will create fields in the adjoint
representation of $USp(n)$, which is equivalent to
the second rank symmetric tensor representation. 
Then, we can guess the fermions of the theory as
in Table \ref{table3}. 
\begin{center}
\refstepcounter{table}
\label{table3}
\footnotesize
{\bf Table \thetable:} Fermions in the $Sp$-type D9-\AD 9 system.
\beq
\begin{array}{c|c|c|c}
\hline\hline
\mbox{string}&\mbox{fermion}
& \mbox{rep. of $USp(n)\times USp(m)$} & \mbox{chirality}\\
\hline
\mbox{\qq string}&\lambda&(\sym,1)&+\\
\mbox{\oo string}&\wt\lambda&(1,\asym)&+\\
\mbox{\qo$\!\!$,\,\oq string}&\psi&(\fnd,\fnd)&-\\
\hline
\end{array}
\nn
\eeq
\end{center}

The condition corresponding to (\ref{mn})
is satisfied if $m-n=32$.
In addition, if we interchange $n$ and $m$,
the identities (\ref{F6-1})--(\ref{F2}) are also satisfied
for the $USp(n)\times USp(m)$ gauge theory, and
all the anomalies are cancelled in the manner
as discussed in \S \ref{GSmech}.

The interpretation in string theory is
as follows. We
fill the spacetime with $n$ D9-branes and $m$ \AD 9-branes,
and we take the $Sp$-type $\Omega$ projection.
It is easy to repeat the analyses of \S\S \ref{Physstat}
and \ref{RRtad}.
For example,
taking into account that $\gamma$ in (\ref{proj}) is $i \J$,
the condition corresponding to (\ref{psisym}) is now
\beq
\J\wt\lambda=-(\J\wt\lambda)^T,
\eeq
implying that the \oo fermions
 $\wt\lambda_{ij}\equiv\J_{ik}\wt\lambda^k{}_j$
belong to the second rank anti-symmetric tensor representation
of $USp(m)$, as expected.

The divergences due to the R-R tadpole are summarized
in Table \ref{table4}.
The total contribution is again cancelled when $m-n=32$.
\begin{center}
\refstepcounter{table}
\label{table4}
\footnotesize
{\bf Table \thetable:} The divergences due to R-R tadpole in $Sp$-type
theory.
\vsx
\beq
\begin{array}{c|c|c|c}
\hline\hline
\mbox{string}&Z_{\RR}^{\rm (C_2)}
&Z_{\RR}^{\rm (M_2)}&Z_{\RR}^{\rm (K_2)}\\
\hline
\mbox{\qq string}&n^2&+2^6n&\\
\mbox{\oo string}&m^2&-2^6m&\\
\mbox{\qo$\!\!$,\,\oq string}&-2mn&&\\
\mbox{closed string}&&&2^{10}\\
\hline
\end{array}
\nn
\eeq
\end{center}

We can also derive this result by counting
R-R charges. The $Sp$-type $\Omega$ projection
can be understood as an effect of an O$9^+$-plane
filling the spacetime.
In ordinary type I string theory, there is an
O$9^-$-plane, which induces
 the $SO$-type 
$\Omega$ projection, and
32 D9-branes are needed to cancel the R-R 10-form charge.
In the $Sp$-type theory,
however, since the sign of the R-R charge
of the O$9^+$-plane is opposite to that of the O$9^-$-plane,
we need 32 \AD9-branes. Therefore,
in the system with $n$ D9-branes and $m$ \AD9-branes,
we must impose the condition $m-n=32$ to cancel the R-R charge.

\section{General formulation of the D9-\AD 9 system}
\label{Gen}
\subsection{Generalization to arbitrary amplitudes}

Adding D9-\AD 9 pairs in type I or type IIB string theory
can be understood as adding additional open strings in the theory.
We have observed from the vacuum amplitudes that
the \oq and \qo strings have the opposite GSO projection, and the
Ramond sector of the \oo strings have the opposite $\Omega$ projection
as the ordinary \qq strings.
This observation should be confirmed in arbitrary
amplitudes, as required by the unitarity of the $S$-matrix.

When we compute the amplitudes in superstring theory,
we must sum over spin structures of
the world-sheet. The spin structures are
characterized by the boundary conditions for the world-sheet
fermions. When we move the fermions around a non-trivial
cycle of the world-sheet, the sign of the fermions can be flipped.
We represent the sign flip by including a cut in the world-sheet.
The cut may end at a boundary of the world-sheet or
a position where a Ramond vertex operator is inserted.
There are holomorphic and anti-holomorphic sectors,
and the spin structures are chosen for each sector.
If the world-sheet has no boundary, the spin structures for
the holomorphic and anti-holomorphic sectors are chosen independently.
However, if the world-sheet has boundaries, holomorphic sectors
and anti-holomorphic sectors are related at the boundaries
by the open string boundary conditions.
Accordingly, if the cuts end at the boundary
in the holomorphic sector, the same holds in the anti-holomorphic sector.
We refer to the boundary, at which odd numbers of cuts end,
as an ``R-R boundary".
In \S\ref{string}, we have assigned an extra minus sign for each
R-R boundary with \AD 9-brane Chan-Paton indices in the vacuum
amplitudes. This prescription can be easily generalized to the case
with arbitrary numbers of boundaries without open string vertex
operators. However, a problem may arise when there
are \oq or \qo string vertex operators at a boundary
of the world-sheet. In this case, the
boundary is broken into pieces with D9-brane
and \AD 9-brane Chan-Paton indices, and
it is ambiguous which sign should be assigned.
In order to resolve this problem, we propose
that an extra minus sign should be assigned for each
endpoint of the cut at the boundary with \AD 9-brane
Chan-Paton indices. The position of the cuts
can be continuously moved without changing any physical
quantities, and hence we must show that
the amplitudes are not changed 
 when we move the cut ending
at the boundary across the \qo or \oq string vertex operators
(Fig.~\ref{fig1}).

\vsm\vsm
\begin{center}
\parbox{5cm}{
\begin{center}
\unitlength=.4mm
\begin{picture}(100,40)(0,0)
\epsfxsize=4cm
\put(0,-10){\epsfbox{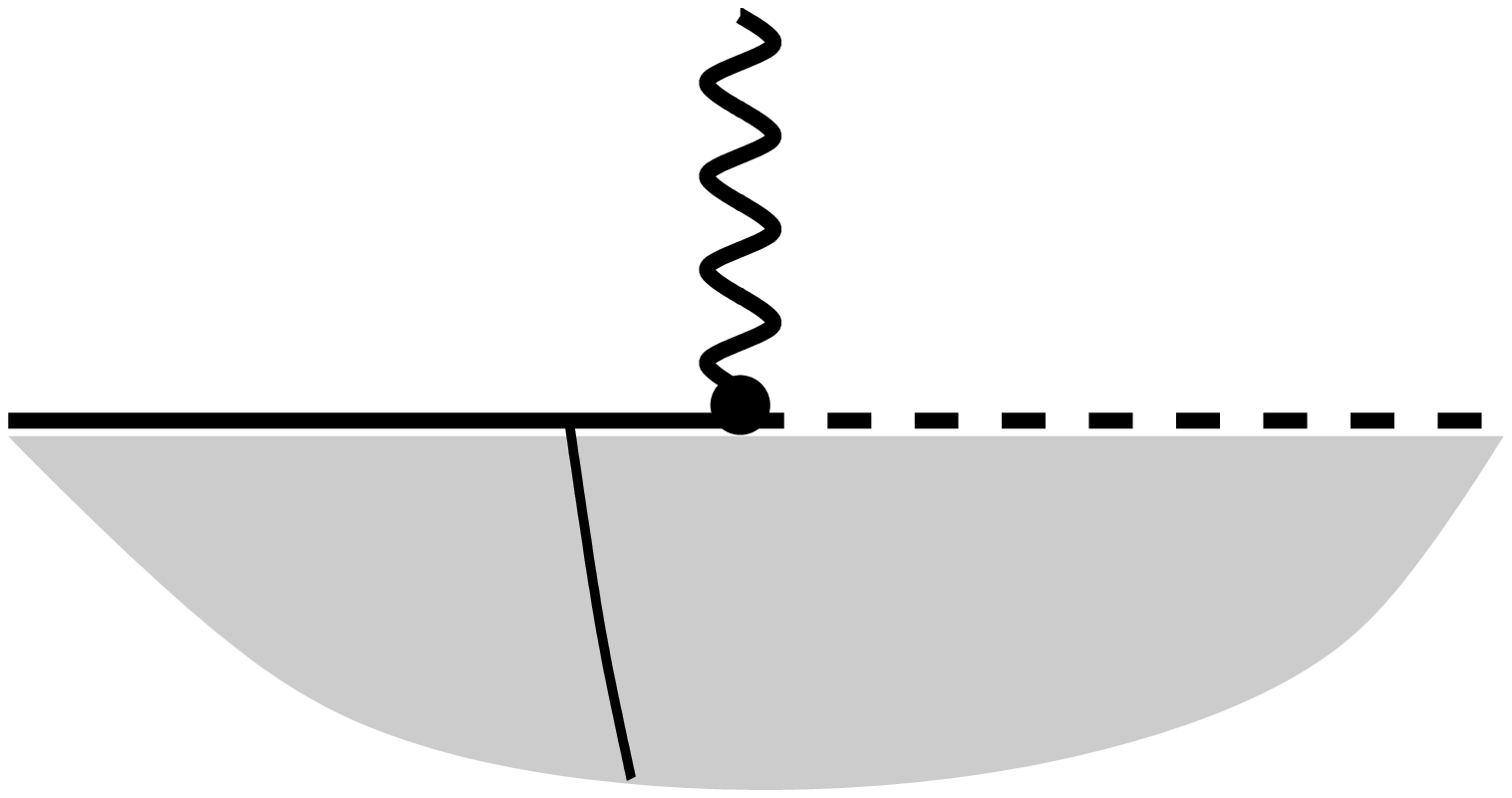}}
\put(32,22){\makebox(0,0){$+1$}}
\end{picture}
\end{center}
}
\begin{picture}(10,10)(0,0)
\put(5,0){\makebox(0,0){$=$}}
\put(5,12){\makebox(0,0){$?$}}
\end{picture}
\parbox{5cm}{
\begin{center}
\unitlength=.4mm
\begin{picture}(100,40)(0,0)
\epsfxsize=4cm
\put(0,-10){\epsfbox{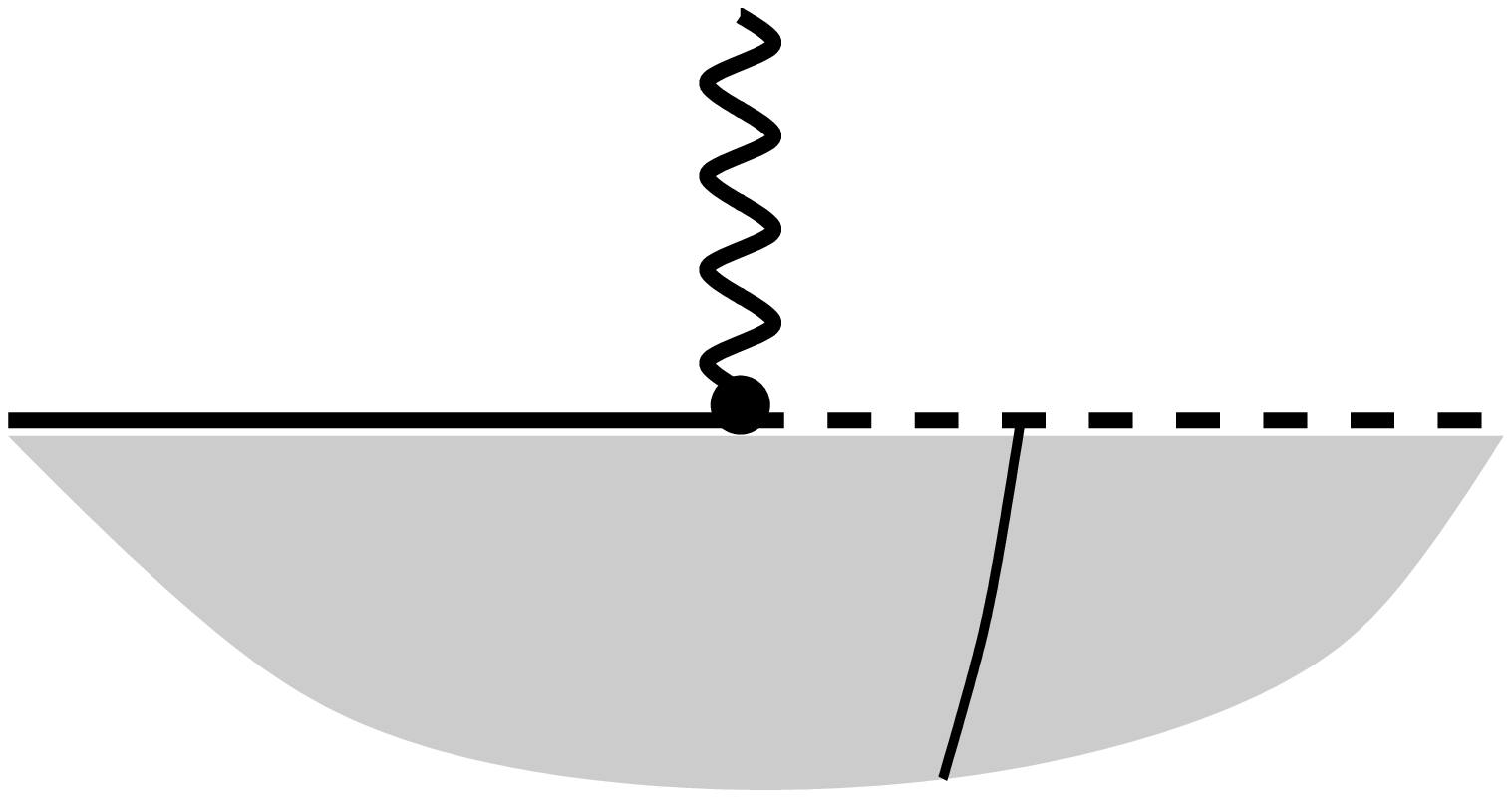}}
\put(66,22){\makebox(0,0){$-1$}}
\end{picture}
\end{center}
}
\vsx\vsx
\refstepcounter{figure}
\label{fig1}
\parbox{14cm}{\footnotesize
{\bf Fig.\thefigure:}
The solid and dashed boundaries are equipped with
D9 and \AD 9-brane Chan-Paton indices, respectively.
We assign an extra minus sign for each endpoint of the cut
at the \AD 9-brane boundary.
}
\end{center}
\vsm

Before solving this problem, let us confirm
that the \qo and \oq strings have the opposite GSO projection.
In the open string channel, making a cut parallel to the spatial
direction of the open string corresponds to the insertion of
the operator
$(-1)^F$ in the operator formalism. If the open string is a \qo or
\oq string, one of the endpoints of the cut is at the
\AD9-boundary, and thus we should assign an extra $-1$ factor.
 Therefore, in our prescription, making a cut parallel
 to the spatial
direction of the \qo or \oq string corresponds to
 the insertion of $-(-1)^F$ in the operator formalism,
as desired (Fig.~\ref{fig2}).
\begin{center}
\parbox{5cm}{
\begin{center}
\unitlength=.4mm
\begin{picture}(100,20)(0,0)
\epsfxsize=3cm
\put(-50,0){\epsfbox{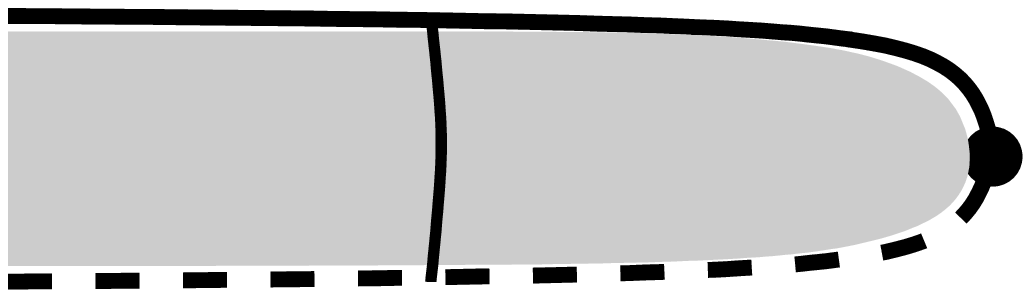}}
\put(-22,-6){\makebox(0,0){$-1$}}
\put(35,10){\makebox(0,0){$\cV_T$}}
\put(50,10){
$=~~-1 \times\left\langle~~\cdots~~ (-1)^F\, \cV_T\right\rangle.$
}
\end{picture}
\end{center}
}

\refstepcounter{figure}
\label{fig2}
{\footnotesize\bf Fig.\thefigure} 
\end{center}
\vsm

Now consider the cut ending at the D9-boundary, as
illustrated in the left-hand side of Figs.~\ref{fig1}
and \ref{fig3}.
This cut can be deformed as in the
right-hand side of Fig.~\ref{fig3}.
\begin{center}
\parbox{5cm}{
\begin{center}
\unitlength=.4mm
\begin{picture}(50,65)(0,0)
\epsfxsize=3cm
\put(0,-10){\epsfbox{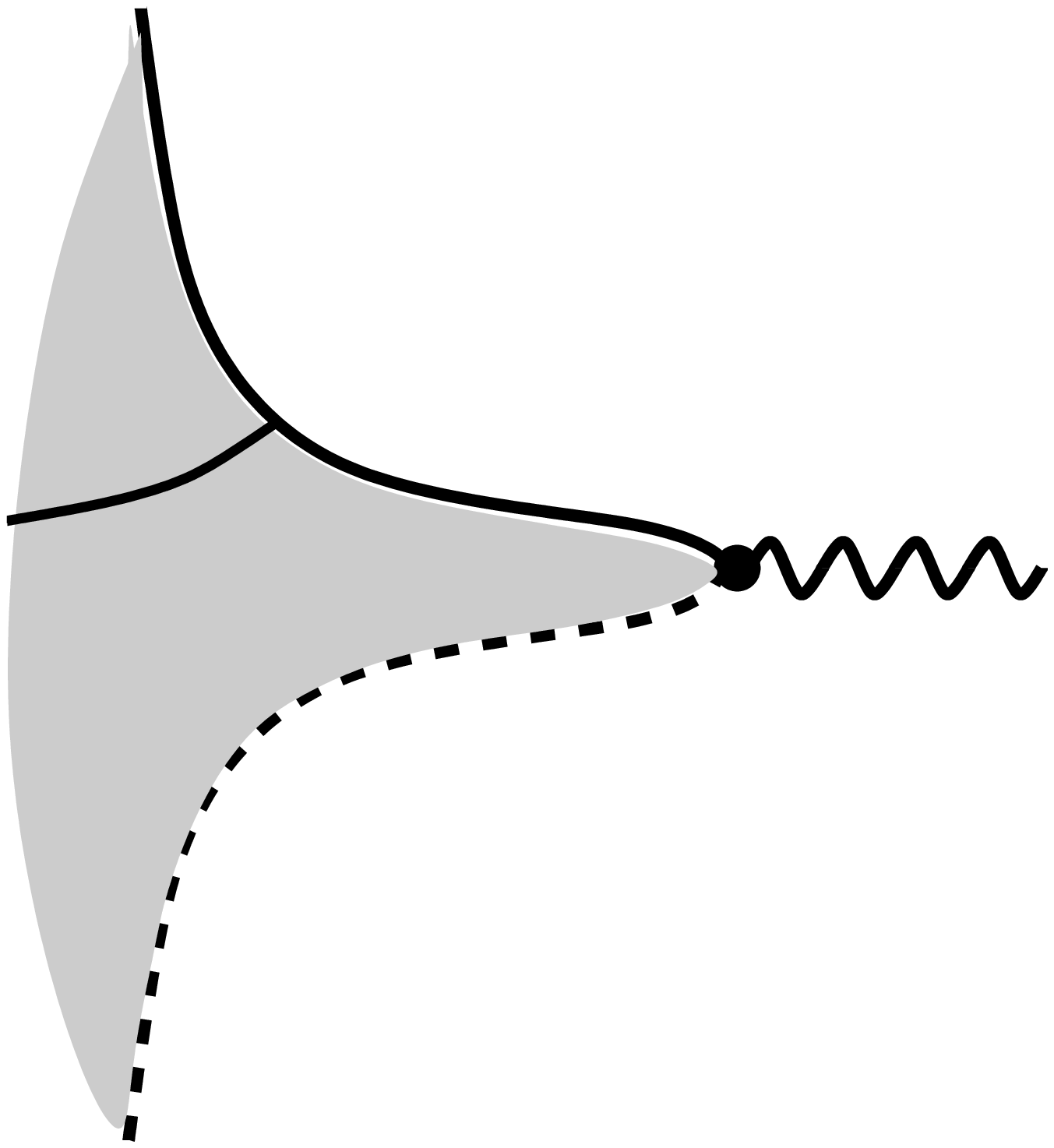}}
\end{picture}
\end{center}
}
\begin{picture}(10,10)(0,0)
\put(10,0){\makebox(0,0){$=$}}
\end{picture}
\parbox{5cm}{
\begin{center}
\unitlength=.4mm
\begin{picture}(50,65)(0,0)
\epsfxsize=3cm
\put(0,-10){\epsfbox{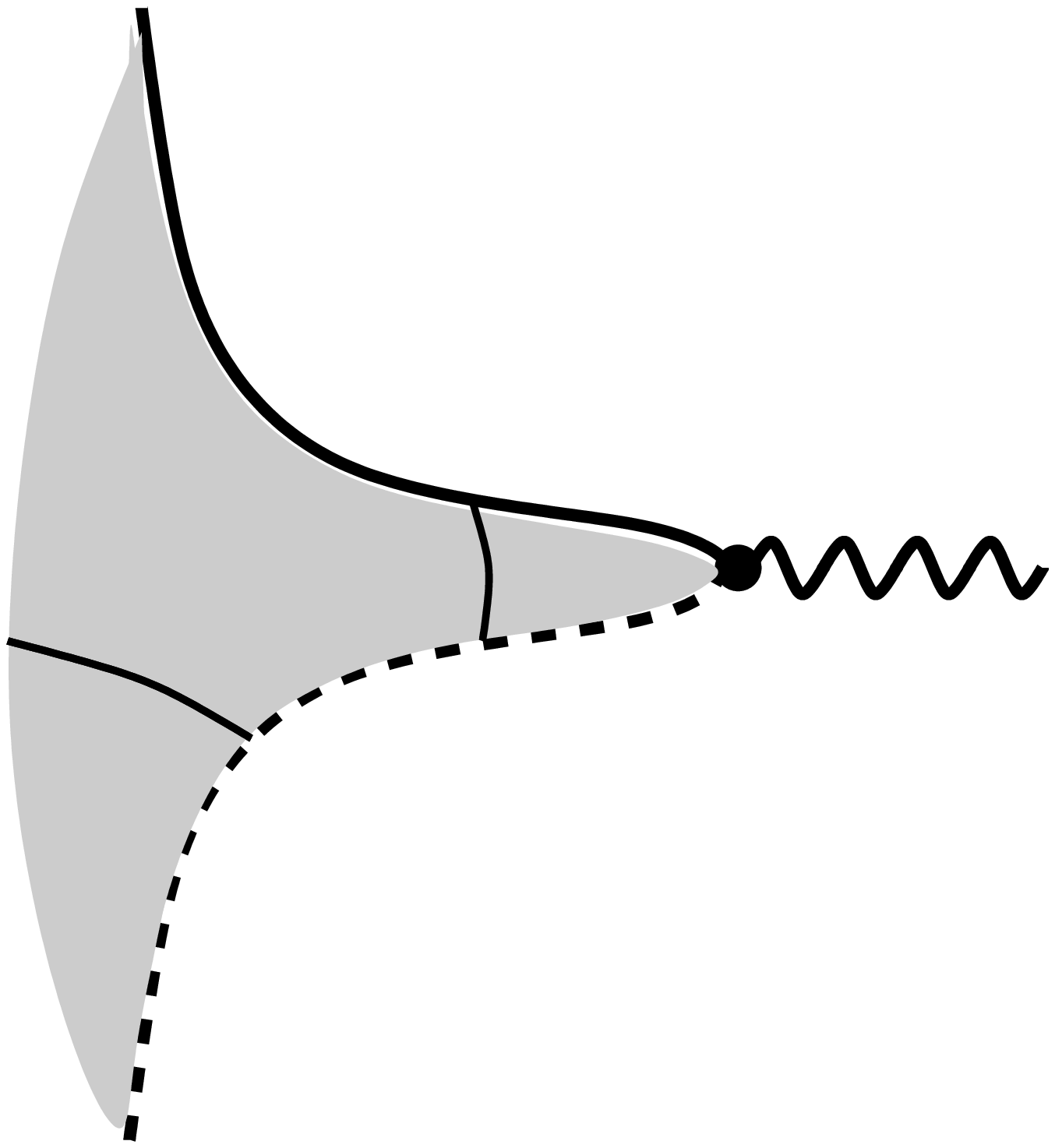}}
\put(37,20){\makebox(0,0){$-1$}}
\put(25,15){\makebox(0,0){$-1$}}
\end{picture}
\end{center}
}

\refstepcounter{figure}
\label{fig3}
{\footnotesize\bf Fig.\thefigure}
\end{center}
\vsm\vsm
\noindent
Then, if the \qo string vertex operator 
in the figure is projected to
satisfy the physical condition $-(-1)^F=1$,
the amplitude is equivalent
to the right-hand side of Fig.~\ref{fig1}.
This is the desired result in order for the
amplitudes to be invariant under continuous deformations
of the cut in the world-sheet.

Next we wish to reconfirm that the Ramond sector of the \oo string
has the opposite $\Omega$ projection. The
open string vertex operator in the Ramond
 sector creates a cut in the world-sheet of
either the holomorphic or anti-holomorphic
sector.\footnote{If necessary, we deform the cut so that it does
not lie along the
boundary of the world-sheet.}
Let us consider the case in which the vertex operator $\cV_R$ creates
a cut in the holomorphic sector.
Then $\Omega\,\cV_R$ will create a cut in the anti-holomorphic sector.
In order to connect the cut consistently, it should end at the boundary,
and hence we need an extra $-1$ factor (Fig.~\ref{fig4}).
Summing up these terms, we have the projection $\Omega=-1$, as desired.

\begin{center}
\parbox{5cm}{
\begin{center}
\unitlength=.4mm
\begin{picture}(100,40)(0,0)
\epsfxsize=3.5cm
\put(-60,32){\epsfbox{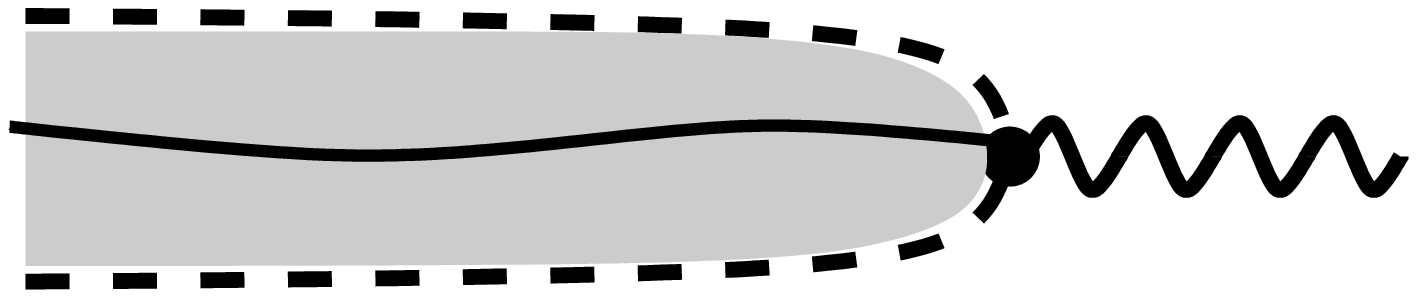}}
\put(40,40){
$=~~\left\langle~~\cdots~~ \, \cV_R\right\rangle.$}
\epsfxsize=3.5cm
\put(-60,-8){\epsfbox{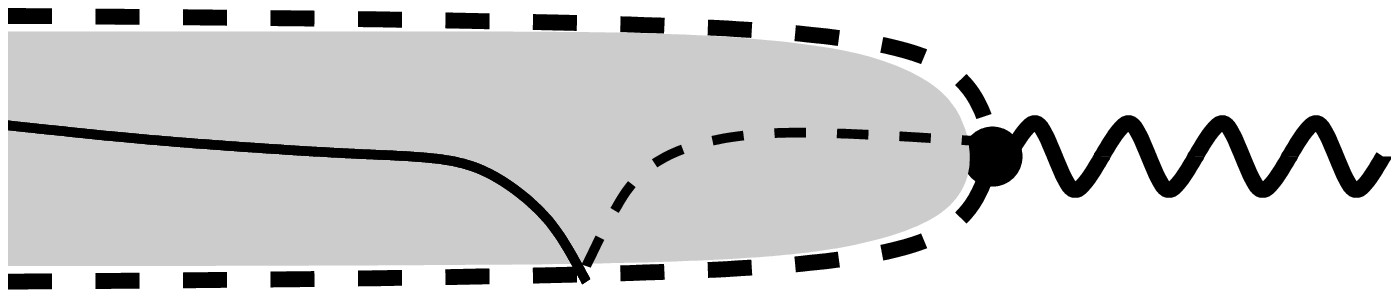}}
\put(-22,-15){\makebox(0,0){$-1$}}
\put(40,0){
$=~~-1 \times\left\langle~~\cdots~~ \Omega\, \cV_R\right\rangle.$}
\end{picture}
\end{center}
}

\vs{3ex}
\refstepcounter{figure}
\label{fig4}
\parbox{14cm}{\footnotesize
{\bf Fig.\thefigure:}
The solid and dashed cuts in the world-sheet 
are the cuts in the holomorphic
and anti-holomorphic sectors, respectively.
}
\end{center}

Note that the cut created by the Ramond sector vertex operator
could be taken in the anti-holomorphic sector. Then the sign of the
amplitude will be flipped. But this is not a problem, since
the overall sign of the amplitude is unphysical.

\subsection{Anomaly cancellations in string theory}

To confirm our prescription,
let us show that the gauge anomaly is cancelled in the 
D9-\AD 9 system.
There are three types of diagrams that contribute to the anomaly,
\cite{GS2,GSW}  planer, non-orientable, and non-planer
orientable diagrams,
as depicted in  Figs.~\ref{planer}--\ref{nonplan}.

\begin{center}
\parbox{4cm}{
\begin{center}
\unitlength=.4mm
\begin{picture}(100,60)(0,0)
\epsfxsize=3.5cm
\put(5,5){\epsfbox{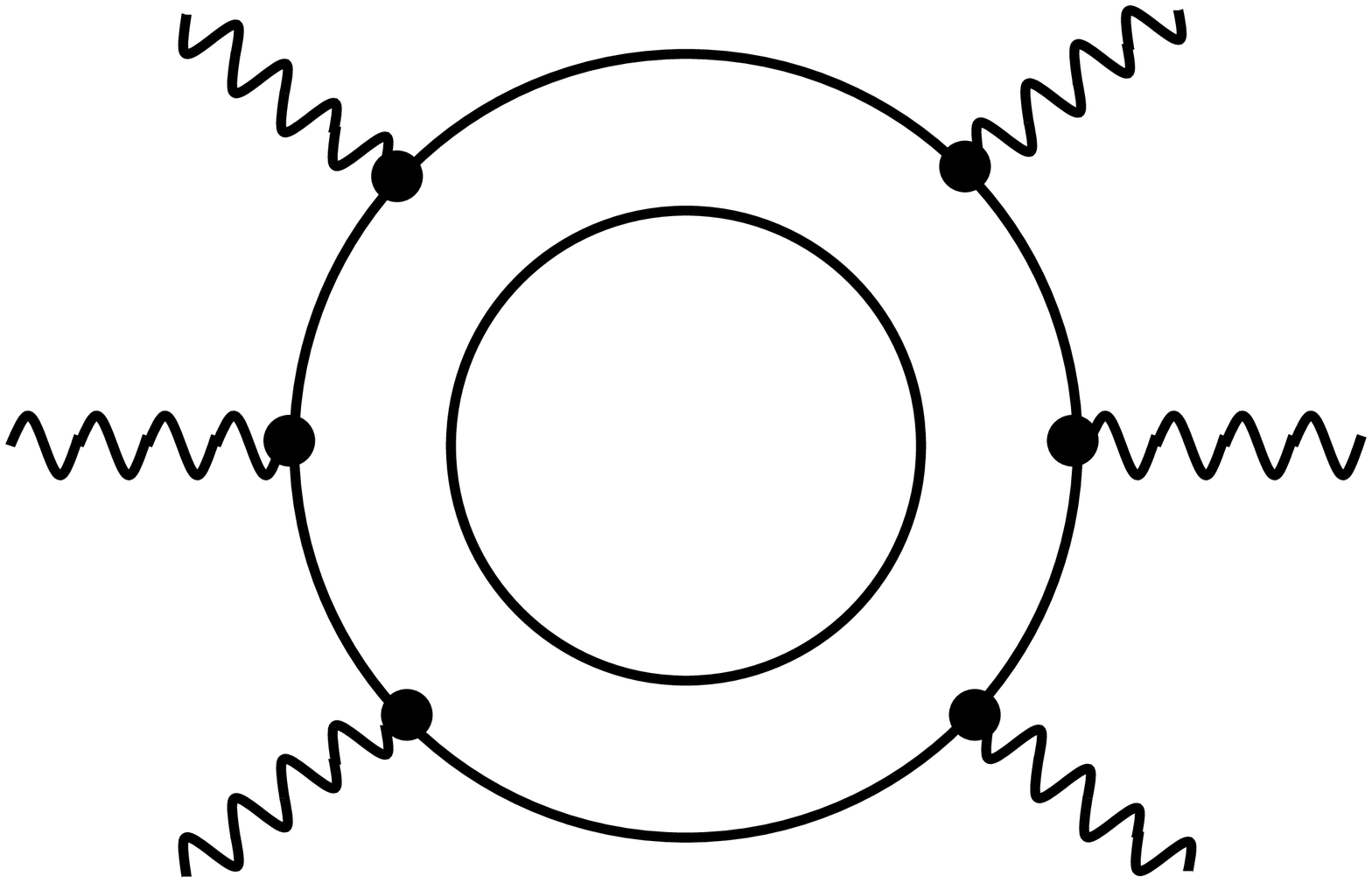}}
\end{picture}
\end{center}
\begin{center}
\refstepcounter{figure}
\label{planer}
\footnotesize
{\bf Fig.\thefigure:} Planer.
\end{center}
}
\parbox{4cm}{
\begin{center}
\unitlength=.4mm
\begin{picture}(100,60)(0,0)
\epsfxsize=3.5cm
\put(5,5){\epsfbox{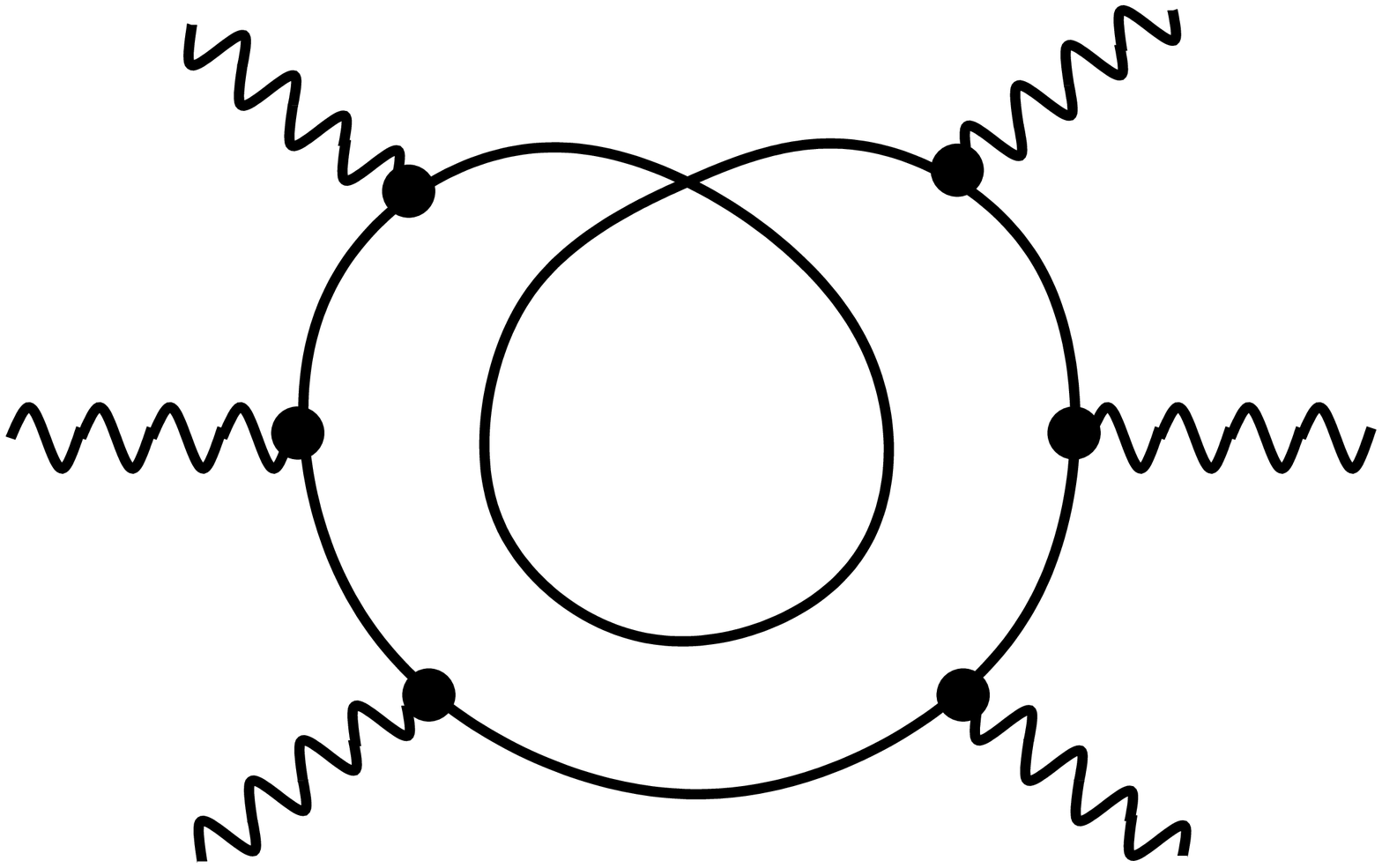}}
\end{picture}
\end{center}
\begin{center}
\refstepcounter{figure}
\footnotesize
{\bf Fig.\thefigure:} Non-orientable.
\end{center}
}
\parbox{4cm}{
\begin{center}
\unitlength=.4mm
\begin{picture}(100,60)(0,0)
\epsfxsize=3.5cm
\put(5,5){\epsfbox{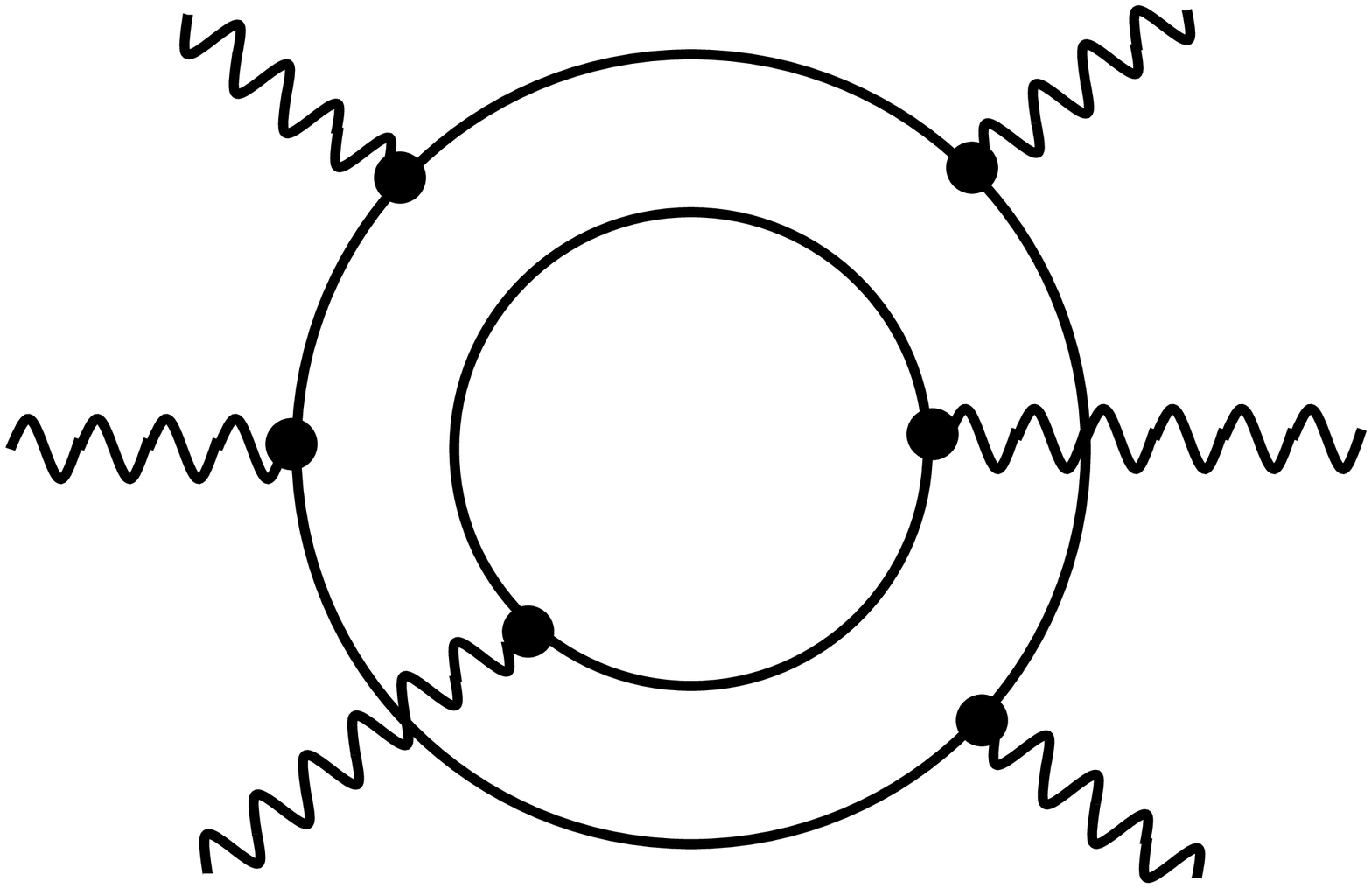}}
\end{picture}
\end{center}
\begin{center}
\refstepcounter{figure}
\label{nonplan}
\footnotesize
{\bf Fig.\thefigure:} Non-planer.
\end{center}
}
\end{center}
\vsm\vsm

We consider these diagrams as open string one-loop diagrams.
Then, only the Ramond sector of the open strings with the $(-1)^F$
insertion  will
contribute to the anomaly, since the
parity-conserving terms are not anomalous.
As explained in Refs.~\citen{GS2} and \citen{GSW}, the non-planer diagrams
are not divergent and do not contribute to the anomaly.
Hence we consider the sum of the contribution from
the planer and non-orientable diagrams.
Let us consider the case that all the external gauge fields
are created by \qq strings.
 The case with \oo gauge fields
can be treated similarly.
The $(-1)^F$ insertion will create a cut in the world-sheet, 
and hence we need an extra $-1$ factor for each \AD 9-brane
boundary in the planer diagrams.
 Thus the contributions from the planer diagrams
are proportional to $n-m$, where $n$ and $m$ are
the numbers of D9 and \AD 9-branes, respectively.
The explicit calculations are
exactly the same as those given in Ref.~\citen{GS2}, except
for this factor. The result is
\beq
{\cal A}=(\,n-m+32\,l\,)\,G,
\eeq
where $G$ is a non-vanishing factor, for which we do not
need detailed structure,
 and $l$ is given as
\beq
l=\left\{
\begin{array}{ccc}
+1,&~~~&USp(n),\\
0,&~~~&U(n),\\
-1,&~~~&SO(n).
\end{array}
\right.
\eeq
As a result, the anomaly is cancelled when $n-m=32\,l$;
namely,
$n=m$ for the type IIB D9-\AD9 system, $n-m=32$ for
the type I D9-\AD9 system, and $m-n=32$ for the D9-\AD9 system
of $USp$-type theory.

\section{Discussion of the $USp(32)$ theory}
\label{USp32}
As the D9-\AD 9 pairs are expected to vanish
after the tachyon condensation,
it is interesting to examine the case with $m=32$
and $n=0$, which is the tachyon-free case.
This theory contains closed strings and open strings
with $USp(32)$ Chan-Paton indices. The formulation is
almost the same as that of the type I $SO(32)$ string theory,
except we take the opposite $\Omega$ projection
for the Ramond sector of the open strings, and
accordingly, we associate an extra minus sign for
each R-R boundary of the world-sheet
in calculating the amplitudes.

Unlike the type I $SO(32)$ theory, the $USp(32)$ theory
does not have a spacetime supersymmetry, since the fermion $\lambda$,
created by the open strings, belongs to the second
rank anti-symmetric tensor representation of
$USp(32)$ and cannot be a supersymmetric partner of the
$USp(32)$ gauge field.  Therefore, there is no reason for
the vacuum energy to be cancelled.
Indeed, there is a divergence in the vacuum amplitude due to
the NS-NS tadpole, that needs to be cancelled by the
Fischler-Susskind mechanism.\cite{FS}
 This induces the cosmological constant term
$\sqrt{-g}\,e^{-\phi}$ in the effective action.

The K-theory analyses given in Ref.~\citen{W} suggest
that there are D1,~3,~4,~5,~9-branes in this theory (Table \ref{table5}).
\begin{center}
\refstepcounter{table}
\label{table5}
\footnotesize
{\bf Table \thetable:} D-branes in the $USp(32)$ theory.
\beq
\begin{array}{c|ccccccccccc}
\hline\hline
k&10&9&8&7&6&5&4&3&2&1&0\\
\hline
KSp(\sR^{k})&&&\sZ&&\sZ_2&\sZ_2&\sZ&&&&\sZ\\
\hline
\mbox{$p$-brane}&-1&0&1&2&3&4&5&6&7&8&9\\
\hline
\end{array}
\nn
\eeq
\end{center}
\vsh
The analyses in Ref.~\citen{GP} show that
D1-branes and D5-branes have $Sp$ and $SO$-type Chan-Paton indices,
respectively. We can also understand this fact using the
isomorphism of K-theory groups $KSp(\R^n)\simeq KO(\R^{n\pm 4})$.
Suppose that we wish to construct lower dimensional D-branes
in the D5-\AD 5 system, as given in Refs.~\citen{Sen} and \citen{W}. In order to
obtain the same D-branes as in Table \ref{table5},
the suitable K-group for the D5-\AD 5 system is $KO(\R^k)$, rather
than $KSp(\R^k)$ (Table \ref{table6}).
\begin{center}
\refstepcounter{table}
\label{table6}
\footnotesize
{\bf Table \thetable:} D-branes in the D5-\AD 5 system.
\beq
\begin{array}{c|ccccccc}
\hline\hline
k&6&5&4&3&2&1&0\\
\hline
KO(\sR^{k})&&&\sZ&&\sZ_2&\sZ_2&\sZ\\
\hline
\mbox{$p$-brane}&-1&0&1&2&3&4&5\\
\hline
\end{array}
\nn
\eeq
\end{center}
Thus, we conclude that the Chan-Paton indices for
D5-branes are of the $SO$-type. A similar argument for D1-branes
shows that D1-branes have $Sp$-type Chan-Paton indices.

The analysis of the spectrum of open strings ending on the
D1-branes is similar to the ordinary type I D-strings,
which is given in Ref.~\citen{PW}. 
Consider $n$ D1-branes lying along the $x^0,~x^9$ direction
in the $USp(32)$ string theory.
Note that $n$ should be even,
since the gauge group on the D1-branes is $USp(n)$.

1-1 strings in the NS sector
create an $USp(n)$ gauge field $A_\mu$ (which can
be gauged away) and eight massless scalar fields
 $X^i$ ($i=1,\cdots,8$), which belong to the second rank
anti-symmetric tensor representation of the gauge
group. World-sheet massless fermions created by
the 1-1 strings in the Ramond sector
 are $S_+^a$ and $S_-^{\hat a}$ ($a,\hat a=1,\cdots,8$).
The subscripts $+$ and $-$ represent the chirality
of the world-sheet Lorentz group $SO(1,1)$,
and the superscripts $a$ and $\hat a$ are indices of the $8_s$ and $8_c$
 spinor representations
of the transverse spacetime Lorentz group $SO(8)$.
$S_+^a$ is a right-moving fermion that belongs to
the adjoint representation of 
$USp(n)$,
while $S_-^{\hat a}$ is a left-moving fermion
that belongs to the second rank anti-symmetric tensor representation
of the gauge group.
1-$\ol 9$ and $\ol 9$-1 strings will create a left moving
fermion $\lambda_-^I$ ($I=1,\cdots,32$), which belongs to
the fundamental representation of $USp(n)$. The  superscript
$I$ comes from the Chan-Paton  indices associated with
32 \AD 9-branes.
\vsx
\begin{center}
\refstepcounter{table}
\label{table7}
\footnotesize{\bf Table \thetable:} Massless spectrum on D1-branes.
\beq
\begin{array}{c|cccc}
\hline\hline
&USp(n)&SO(8)&USp(32)\\
\hline
A_\mu &\sym&1&1\\
X^i&\asym&8_v&1\\
S_+^a&\sym&8_s&1\\
S_-^{\hat a}&\asym&8_c&1\\
\lambda_-^I&\fnd&1&\fnd\\
\hline
\end{array}
\nn
\eeq
\end{center}

Let us consider the minimal case $n=2$.
In this case, the gauge group is $USp(2)\simeq SU(2)$,
and the second rank anti-symmetric tensor representation
is a singlet.
The action is
\beq
S=\int\! d^2\sigma\,\left(\, -\frac{1}{4}(F_{\mu\nu})^2+
\half(\del_\mu X^i)^2+
S^a_+[D_-,S^a_+]+S^{\hat a}_-\del_+S^{\hat a}_-
+\lambda^I_-D_+\lambda^I_-
\,\right),
\eeq
where we have defined $D_\pm=\del_\pm+i A_\pm$.
In analogy to the argument in
 S-duality for type I and heterotic
string theory, \cite{PW} it has been
suggested that this action is
the action of the fundamental string in the heterotic
version of $USp(32)$ string theory.
It would be interesting to investigate
the detailed structure of this theory.

\section*{Acknowledgments}
The author  would like to thank our colleagues at Kyoto University
and YITP for valuable discussions and encouragement. 
He is  especially grateful to H.~Kunitomo, Y.~Imamura,
 N.~Sasakura and Y.~Michishita
for useful discussions. This work is supported in part
by a Grant-in-Aid for JSPS fellows.

\end{document}